\documentclass{PoS}

\title{Hints for an axion-like particle from PKS 1222+216?}

\ShortTitle{Hints for an axion-like particle from PKS 1222+216?}

\author{\speaker{Giorgio Galanti}%
        \\
        Universit\`a dell'Insubria, Como, Italy\\
        E-mail: \email{gam.galanti@gmail.com}}

\author{Marco Roncadelli\\
        INFN, Pavia, Italy\\
        E-mail: \email{marco.roncadelli@pv.infn.it}}

\author{Fabrizio Tavecchio\\
        Osservatorio di Brera-INAF, Milano, Italy\\
        E-mail: \email{fabrizio.tavecchio@brera.inaf.it}}

\author{Giacomo Bonnoli\\
        Osservatorio di Brera-INAF, Milano, Italy\\
        E-mail: \email{giacomo.bonnoli@brera.inaf.it}}

\abstract{Flat spectrum radio quasars (FSRQs) are a particular class of blazars rich of optical/ultraviolet photons inside the broad line region (BLR), necessarily implying a huge optical depth for $\gamma$ rays above $20 \, {\rm GeV}$. As a consequence, photons with energy above such a threshold should not be emitted. However, photons in the energy range $70 - 400 \, {\rm GeV}$ have been observed by MAGIC from the FSRQ PKS 1222+216. Several astrophysical explanations exist in the literature, but they are all {\it ad hoc}, namely devised only for that specific purpose. We show that such a surprising discovery can be explained within standard blazar models by adding the new possibility that photons oscillate into axion-like particles (ALPs) and vice-versa inside the source. Through the photon-ALP oscillation mechanism a sizable fraction of very-high energy (VHE) photons can escape absorption from the BLR in a similar fashion as they largely avoid absorption from the extragalactic background light (EBL) in the intergalactic space. Actually, we show that not only are VHE photon indeed emitted, but also that their spectral energy distribution (SED) is such that they lie on the same Compton peak to which also lower energy photons simultaneously detected by {\it Fermi}/LAT belong.}

\FullConference{Science with the New Generation of High Energy Gamma-ray experiments, 10th Workshop \\
                 04-06 June 2014\\
                 Lisbon - Portugal}

\begin{document}

\section{Introduction}

The Standard Model (SM) of strong, weak and electromagnetic interactions describes all experimental data up to about the Fermi scale $G_F^{- 1/2} \simeq 250 \, {\rm GeV}$ with striking precision, and the discovery of the Higgs boson has further supported its success. However, it is nowadays well known that the SM cannot be considered as the ultimate theory of fundamental processes, but should merely be regarded as the low-energy manifestation of some more fundamental theory which includes gravity and explains both the non-baryonic dark matter and the dark energy. Plenty of proposals for a satisfactory extension of the SM exist in the literature, chiefly among them superstring theories. Quite remarkably, many different approaches generically predict the existence of axion-like particles (ALPs). ALPs are very light pseudo-scalar spin-zero bosons and give rise to photon-ALP oscillations in the presence of an external magnetic field.

While ALPs are unobservable in present-day accelerator experiments, their existence can likely show up in {\it blazar} observations (more about this, later) in the very-high-energy (VHE) band ($100 \, {\rm GeV} < E < 100 \, {\rm TeV}$) with the presently operating Imaging Atmospheric Cherenkov Telescopes (IACTs) H.E.S.S., MAGIC and VERITAS, even though a much better chance in this respect is offered by the planned Cherenkov Telescope Array (CTA).

Some hints of the ALP existence have been reported: (i) they provide a solution of the so-called {\it pair-production anomaly}~\cite{horns}, (ii) they can explain the observed low value of the extragalactic magnetic field~\cite{finke}, and (iii) they offer a natural solution to the {\it cosmic opacity problem} by substantially preventing photon absorption from the extragalactic background light 
(EBL)~\cite{dgr}, thereby also considerably extending the $\gamma$-ray horizon.

Our aim is to present an additional hint supporting the existence of ALPs, namely the VHE emission by flat spectrum radio quasars (FSRQs) at energies well above $20 \, {\rm GeV}$, which is absolutely prevented by conventional physics. Basically, we show that the surprising $\gamma$-ray detection of PKS 1222+216 by MAGIC in the energy range $70 - 400 \, {\rm GeV}$ is explained within the standard FSRQ models provided that we allow for the new possibility of photon-ALP oscillations inside the source~\cite{trgb}. The result reported here can also be applied with minor modifications to the other FSRQs observed in the VHE band, like for instance 3C 279.

\section{Axion-like particles (ALPs)}

ALPs (for a review, see~\cite{alprev}) are a generalization of the axion, the pseudo-Goldstone boson arising from the breakdown of the global Peccei-Quinn symmetry $U(1)_{\rm PC}$ proposed to solve the well known strong {\it CP} problem. In contrast with the axion, for an ALP -- to be denoted by $a$ -- the mass $m$ and the $a \gamma \gamma$ coupling constant $1/M$ are {\it unrelated} parameters. In addition, {\it only} the ALP interaction with two-photons is considered -- mainly to make the analysis as much as model-independent as possible -- and so ALPs are described by the Lagrangian 
\begin{equation}
\label{t1}
{\cal L}^0_{\rm ALP} = \frac{1}{2} \, \partial^{\mu} a \, \partial_{\mu} a - \frac{1}{2} \, m^2 \, a^2 + \frac{1}{M} \, {\bf E} \cdot {\bf B} \, a~,
\end{equation}
where for our purposes $\bf B$ is an {\it external} magnetic field while $\bf E$ is the electric field of a propagating photon. Only the CAST experiment gives a robust bound on $M$ which reads $M > 1.14 \cdot 10^{10} \, {\rm GeV}$ for $m < 0.02 \, {\rm eV}$~\cite{cast}. In order to complete our scenario it is necessary to include the photon QED one-loop vacuum polarization in the presence of an external 
${\bf B}$, represented by the Heisenberg-Euler-Weisskopf (HEW) effective Lagrangian~\cite{HEW} 
\begin{equation}
\label{t1q}
{\cal L}_{\rm HEW} = \frac{2 \alpha^2}{45 m_e^4} \, \left[ \bigl({\bf E}^2 - {\bf B}^2 \bigr)^2 + 7 \bigl({\bf E} \cdot {\bf B} \bigr)^2 \right]~,
\end{equation}
where $\alpha$ is the fine-structure constant and $m_e$ is the electron mass. We emphasize that ${\cal L}_{\rm HEW}$ becomes of paramount importance when dealing with external magnetic fields of ${\mathcal O}(1 \, {\rm G})$ and energies above ${\mathcal O}(100 \, {\rm GeV})$, as in the case of interest. So, we will be concerned throughout with the full Lagrangian ${\cal L}_{\rm ALP} = {\cal L}^0_{\rm ALP} + {\cal L}_{\rm HEW}$.

As it is evident from the last term in Eq. (\ref{t1}) -- which is responsible for the $a \gamma \gamma$ coupling -- only the component of $\bf B$ transverse to the photon momentum and parallel to its polarization couples to $a$: it will be henceforth denoted by ${\bf B}_T$. Moreover, in the presence of an external magnetic field the $a \gamma \gamma$ coupling produces a mismatch between the interaction eigenstates and the propagation eigenstates, thereby giving rise to the phenomenon of photon-ALP oscillations. The present situation is similar to the oscillations of massive neutrinos of different flavors, apart from the difference that here an external ${\bf B}$ field is necessary in order to compensate for the spin mismatch between photons and ALPs.

\section{Flat spectrum radio quasars (FSRQs)}

Active galactic nuclei (AGN) are basically supermassive black holes (SMBHs) located at the centre of bright elliptical galaxies and accreting matter from the surrounding, which -- before disappearing into the SMBH -- heats up tremendously and consequently radiates an enormous amount of {\it thermal} energy. Roughly $10 \, \%$ of the AGN possess two opposite relativistic and highly collimated jets orthogonal to an accretion disk. Whenever one of the jets happens to be directed towards the Earth, the AGN is called {\it blazar}. Electrons accelerated in the jet are a source of {\it non-thermal} radiation, which spans the entire electromagnetic spectrum. The corresponding spectral energy distribution (SED) is characterized by two humps. The first one peaks somewhere between the infrared and the x-ray band and is due to the synchrotron emission of relativistic electrons in the jet. The second peak lies in the $\gamma$-ray band, but its origin is debated. Two mechanisms have been proposed for its origin: one leptonic and the other hadronic. In the leptonic case~\cite{lept} the peak is due to the inverse Compton scattering off the same electrons responsible for the synchrotron peak (with a possible contribution from external photons), while in the hadronic mechanism~\cite{hadr} the considered peak is due to reactions involving relativistic hadrons with neutral and charged pions decaying into $\gamma$-rays and neutrinos, respectively.

Blazars are further divided into two broad classes: BL Lac objects (BL Lacs) and flat spectrum radio quasars (FSRQs). BL Lacs show very weak or even no emission lines in their spectra, hence they are believed to be poor of soft photons. On the contrary, FSRQs show intense broad emission lines arising from the existence of photo-ionized clouds rich of ultraviolet photons and rapidly rotating around the central SMBH, giving rise to the so-called {\it broad line region} (BLR) at about $10^{18} \, {\rm cm}$ from the centre. Because of the very high density of ultraviolet photons in the BLR, photons with $E > 20 \, {\rm GeV}$ -- which are produced {\it before} the BLR along the jet by about two orders of magnitudes -- are absorbed due to the process $\gamma \gamma \to e^+ e^-$ when the jet crosses the BLR. As a result, FSRQs have an optical depth for VHE photons which is huge so that photons with energy $E > 20 \, {\rm GeV}$ should be totally unobservable, as it is shown by optical depth depicted by the blue dashed line in the left panel of Fig.~\ref{results}.

\section{A model for the FSRQ PKS 1222+216}

While monitored by {\it Fermi}/LAT in the energy range $0.3 - 3 \, {\rm GeV}$, FSRQ PKS 1222+216 was also detected by MAGIC in the energy range $70 - 400 \, {\rm GeV}$. Further, MAGIC found a flux doubling in about 10 minutes, which implies that the VHE emitting region must have a typical size of about $10^{14} \, {\rm cm}$. So, two challenges arise at once. Why does such a FSRQ emit in the VHE range, contrary to any conventional expectation? Why is the VHE emitting region so small and yet so powerful? 

A few astrophysical solutions have been put forward. For instance, one is the existence of collimated beams of neutral particles produced in the inner jet, which propagate unimpeded through the BLR and produce leptons outside it~\cite{dermer}. Others  amount to locate one or more VHE emission regions outside the BLR~\cite{nalewajko}. However, all astrophysical proposals of this kind appear totally {\it ad hoc}, in the sense that they have been devised {\it only} to explain the observation in question.

Below, we explore a novel possibility which naturally solves the problem and can also be applied to all other FSRQs with minor modifications: VHE photons are produced in the central region as in the standard FSRQ models, but we allow for photon-ALP oscillations to take place in the magnetic field of the source jet and of the host galaxy. Therefore, we expect most of the photons to become ALPs {\it before} reaching the BLR -- thereby substantially avoiding the BLR absorption -- and to reconvert back into photons {\it after} the BLR, either in the magnetic field of the outer jet or of the host galaxy, in such a way that they are ultimately detected as ordinary photons. 

Our job is now clear. Within the scenario sketched above, we should not only explain the MAGIC observations but we also have to provide a realistic model which gives rise to a SED that {\it simultaneously} fits both the {\it Fermi}/LAT and EBL-deabsorbed MAGIC data. Roughly speaking, both data sets should stay on the {\it same} inverse Compton peak. We shall denote by $d$ the distance along the jet measured from the central SMBH, and specifically by $d_{\rm PR}$ and $d_{\rm BLR}$ the distance of the VHE $\gamma$-ray production region and of the BLR, respectively. Finally, we stress again that $d_{\rm PR} \ll d_{\rm BLR}$.

So, our goal is to ultimately evaluate the photon survival probability $P_{\gamma \to \gamma} (E)$ along the path of the photon/ALP beam from the $\gamma$-ray production region of the FSRQ up to the Earth. We assume a disk luminosity $L_{\rm D} \simeq 1.5 \cdot 10^{46} \, {\rm erg \, s^{-1}}$ and we work for fixed values of $M$ and $m$ (more about this, later). Correspondingly, we  divide the path in question into four parts. 

\begin{figure}[h]
\centerline{\includegraphics[width=1.07\textwidth]{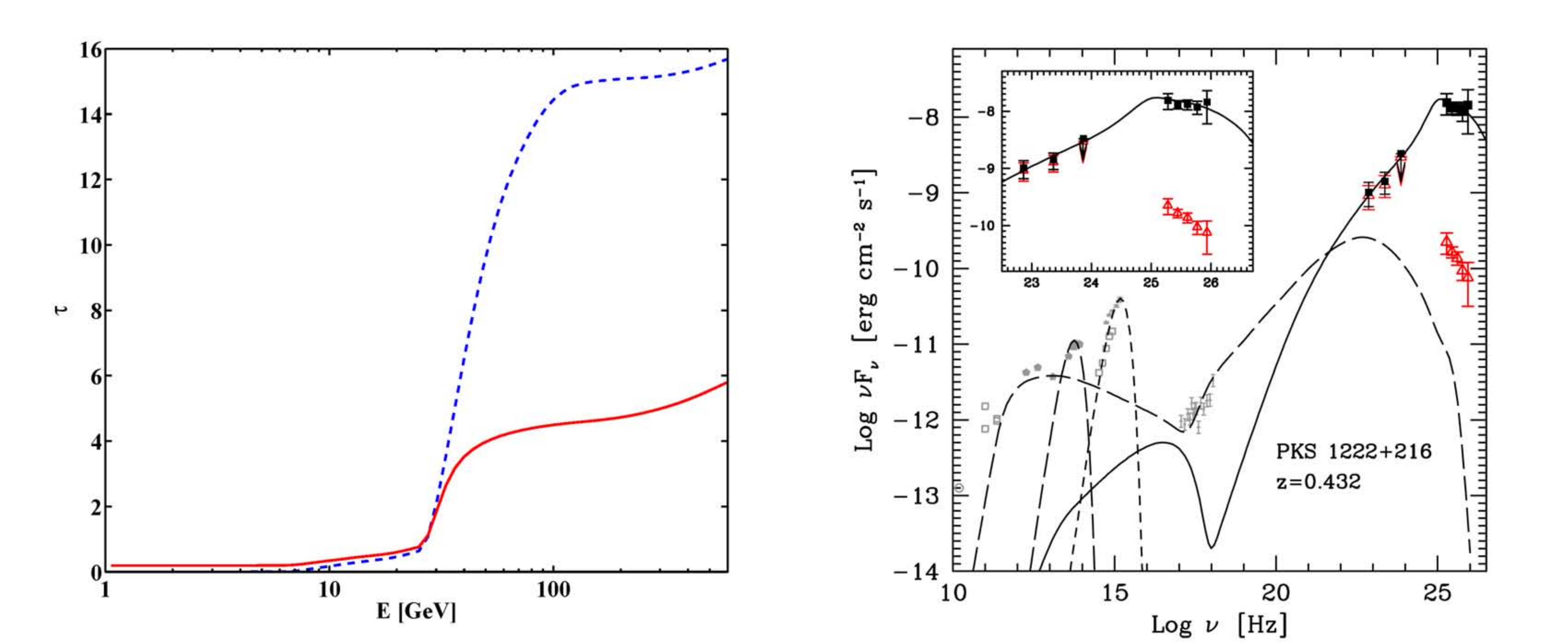}}
\caption{{\it Left panel}: Optical depth as a function of energy computed according to conventional physics (blue dashed line), and with photon-ALP oscillations at work in our best case (red solid line). {\it Right panel}: Red triangles are the {\it Fermi}/LAT and MAGIC EBL-deabsorbed data, black filled squares are the same data once further corrected for the photon-ALP oscillation effect in our best case (see text), and the solid black line is the SED of our model (all non-solid lines should be ignored).}\label{Fig:B}
\label{results}
\end{figure}

\begin{itemize} 

\item {\it Photon-ALP oscillations before and in the BLR} -- This part extends from $d_{\rm PR} \simeq 10^{16} \, {\rm cm}$ to $d_{\rm BLR} \simeq 10^{18} \, {\rm cm}$. Because of lack of information, we assume that here the magnetic field is constant and equal to $B \simeq 0.2 \, {\rm G}$. Still, due to unknown orientation of ${\bf B}$, we suppose that on average ${\bf B}$ is at $45^{\circ}$ with respect to the beam direction, which amounts to take $B_T \simeq 0.14 \, {\rm G}$. Next, under the realistic assumption that the BLR clouds are confined by both thermal and magnetic pressure, we find that the electron number density in the BLR is $n_e \simeq 10^{4} \, {\rm cm^{-3}}$. Finally, VHE beam photons scatter off optical/ultraviolet photons in the BLR through the process $\gamma \gamma \to e^+ e^-$, which is responsible for the huge optical depth for VHE photons in the absence of photon-ALP oscillations (plotted by the blue dashed line in the left panel of Fig.~\ref{results}) and produces a photon absorption which can be properly taken into account in the framework outlined in the second Section.

\item {\it Photon-ALP oscillations in the outer jet} -- Beyond the BLR the photon absorption and the electron number density become negligible. In this region the magnetic field has both a toroidal $B_{\rm tor} \propto 1/d$ and a poloidal $B_{\rm pol} \propto 1/d^2$ component. Because here $d$ is large, the toroidal component dominates. Actually, the outer boundary of this region is set by the condition that $B_{\rm tor} = B_{\rm host}$, where $B_{\rm host}$ denotes the magnetic field in the host elliptical galaxy. 

\item {\it Photon-ALP oscillations in the host galaxy} -- Again, the photon absorption and the electron number density are negligible.  Supernova explosions and stellar motion give rise to a turbulent random magnetic field modeled by a domain-like structure, with strength $B_{{\rm host}} \simeq 5 \, \mu {\rm G}$ and domain length $L_{{\rm coher}} \simeq 150 \, {\rm pc}$~\cite{mossshukurov}. This region extends up at the galaxy border, which can be taken to be the Holmberg radius. 

\item {\it Photon-ALP oscillations in the extragalactic space} -- In this region photon absorption still becomes important due to the scattering of VHE beam photons with EBL photons. A very detailed treatment of photon-ALP oscillations in extragalactic space brought about by an intergalactic magnetic field in the $0.1 - 1 \, {\rm nG}$ range can be found in~\cite{dgr}. However, it turns out that what happens in this region is of minor importance since its affects only very slightly our final results.

\end{itemize}
Because of lack of space we cannot report here the details of the calculation of the propagation process and we refer the reader to our original paper~\cite{trgb}. By a trial-and-error procedure we have found that our best case corresponds to the choice $M = 7 \cdot 10^{10} \, {\rm GeV}$ and $m < 10^{- 10} \, {\rm eV}$.

Let us next address the SED. A model which reproduces {\it all} observations of PKS 1222+216 consists in two blobs located in the central region: the larger one responsible for the emission from the infrared to the x rays, and the smaller one which gives rise to the $\gamma$ rays (see~\cite{trgb} for details).

Our findings are illustrated in Fig.~\ref{results}.

\section{Conclusions}

Including photon-ALP oscillation into standard FSRQ models (like the Synchro-Self-Compton one) we have been able to solve the conundrum posed by the MAGIC observations of PKS 1222+216. Because this source has been simultaneously detected also by {\it Fermi}/LAT, our scenario is logically consistent only if in addition our SED fits both data sets. And we have shown that this is indeed the case for a realistic two-blob model. It looks tantalizing that basically the same choice of the free parameters $M$ and $m$ adopted here are consistent with those that provide the other hints of the existence of ALPs mentioned in the Introduction. Finally, we stress that ALPs with same properties considered here can be discovered by the presently operating IACTs like H.E.S.S., MAGIC and VERITAS, and more likely with the planned observatories CTA, HAWK and HiSCORE. A very remarkable fact is that the same goal can be independently achieved by the laboratory experiments ALPS II at DESY and IAXO, as well as with those based on the techniques discussed in~\cite{avignone}.

\end{document}